\documentstyle[art12,bezier]{article}

\def \sect #1 {\setcounter{equation} 0\section{#1}}
\def \be  {\begin{equation}}
\def \ee  {\end{equation}}
\def \ba  {\begin{eqnarray}}
\def \ea  {\end{eqnarray}}
\def \baa {\begin{eqnarray*}}
\def \eaa {\end{eqnarray*}}
\def \bb  {\begin{thebibliography}}
\def \eb  {\end{thebibliography}}
\newcommand \ci [1] {\cite{#1}}
\newcommand \bi [1] {\bibitem{#1}}
\def \lab #1 {\label{#1}}
\newcommand\re[1]{(\ref{#1})}

\def \qqqquad {\qquad\qquad}
\newcommand\lr[1]{{\left({#1}\right)}}

\def \vev #1 {\langle{#1}\rangle}
\def \VEV #1 {\left\langle{#1}\right\rangle}
\def \ket #1 {|{#1}\rangle}
\def \bra #1 {\langle {#1}|}
\def \e {\mbox{e}}
\def \CO {{\cal O}}
\def \CP {{\cal P}}
\def \CT {{\cal T}}

\def \fracs #1#2 {\mbox{\small $\frac{#1}{#2}$}}
\def \partder #1 {{\partial \over\partial #1}}
\def \bin #1#2 {{\left({#1}\atop{#2}\right)}}
\def \as {\relax\ifmmode\alpha_s\else{$\alpha_s${ }}\fi}
\def \alpi {\frac \as \pi}
\def \al #1 {\frac {\as({#1})}{\pi} }
\def \ds #1 {\ooalign{$\hfil/\hfil$\crcr$#1$}}

\def \QCD {\mbox{{\tiny QCD}}}

\begin{document}

\def\thefootnote{\fnsymbol{footnote}}
\thispagestyle{empty}
\hfill\parbox{35mm}{{\sc ITP--SB--93--73}\par
                         hep-ph/9311294  \par
                         October, 1993}
\vspace*{45mm}
\begin{center}
{\LARGE On near forward high energy scattering in QCD}
\par\vspace*{22mm}\par
{\large Gregory~P.~Korchemsky}%
\footnote{On leave from the Laboratory of Theoretical Physics,
          JINR, Dubna, Russia}
\footnote{{\tt e-mail:~korchems@max.physics.sunysb.edu}}
\par\bigskip\par\medskip
{\em Institute for Theoretical Physics, \par
State University of New York at Stony Brook, \par
Stony Brook, New York 11794 -- 3840}
\end{center}
\vspace*{20mm}

\begin{abstract}
We consider elastic quark-quark scattering at high energy and fixed
transferred momentum. Performing factorization of soft gluon exchanges
into Wilson lines vacuum expectation values and studying their properties,
we find that the asymptotics of the scattering amplitude is controlled
by the renormalization properties of the so called cross singularities
of Wilson loops. Using this fact, we evaluate the scattering amplitude
and show that its asymptotics is determined by the properties of the
$2\times 2$ matrix of anomalous dimensions which appears after one
renormalizes the cross singularities of Wilson loops. A generalization
to the case of quark-antiquark and gluon-gluon elastic scattering is
discussed.
\end{abstract}

\newpage
\def\thefootnote{\arabic{footnote}}
\setcounter{footnote} 0

\section{Introduction}

Investigation of the hadron-hadron interaction at high energy $s$ and
fixed transferred momentum $t$ is one of the longstanding problems
in QCD \ci{book1,book2}. The first calculations of the elastic near
forward parton-parton scattering amplitudes to the lowest orders of
perturbation theory (PT) in the leading logarithmic approximation
revealed novel features of nonabelian gauge theories \ci{PT}. It was
found in QCD that gluons, in contrast with photons in QED \ci{QED},
are reggeized and their contribution to the scattering amplitude has
the asymptotics $\sim s^{\alpha(t)}$ with the Regge trajectory
$\alpha(t)$ known to the lowest order of PT. Contributions of individual
Feynman diagrams to the scattering amplitude obey a remarkable property.
In the center of mass frame of the incoming particles the $s-$dependence
of diagrams comes from integration over longitudinal components of the
exchanged gluon momenta whereas the $t-$dependence is given by
two-dimensional integrals over transverse components of gluon momenta.
It was proposed \ci{2-dim} to explore this property in order to describe
the asymptotics of the scattering amplitudes within the framework of
some effective two-dimensional field theory. Lipatov succeeded to resum
the so called $i\pi-$corrections using explicit expressions for the
$S-$matrix of exactly solvable two-dimensional field theories
\ci{book1,i-pi}. Most recently, another two-dimensional field theory
was proposed to describe high-energy behavior in QCD \ci{rec}. Although
this theory reproduces the known expressions for the amplitude to the
lowest orders of PT in the leading logarithmic approximation, it is very
hard to solve the model exactly to find the asymptotics of the scattering
amplitude to all orders of PT.

The asymptotics of the parton-parton scattering amplitude was found in
the leading logarithmic approximation using the ``evolution equation''
approach \ci{book1,i-pi,evol}. In this approach, one uses the results of
one-loop calculations of the scattering amplitude in order to resum the
leading logarithmic corrections to all orders of PT. However, in spite of
considerable progress, we don't have a regular way to resum and control
nonleading logarithmic corrections to the scattering amplitude. In the
present paper we propose a new approach which enables us to take into
account systematically these nonleading corrections.

\section{Factorization of soft gluons}

Let us consider near forward hadron-hadron scattering with large energy
$s$ and fixed transferred momenta $t$ which is chosen to be much larger
than the QCD scale $\Lambda_{\QCD}$. In this kinematics, interaction
between hadrons occurs at short distances and we may apply the parton model
to relate the asymptotics of the process to properties of the parton-parton
scattering. We start by considering in detail a near forward elastic
quark-quark scattering and generalize consideration to quark-antiquark
and gluon-gluon scattering at the end of the paper.

Let $p_1, p_2$ and $p'_1, p'_2$ be the momenta of incoming and outgoing
scattered quarks, respectively. We choose all these momenta to be on-shell
$p_1^2=p_2^2={p'}_1^2={p'}_2^2=m^2$ with $m$ being quark mass and define
the kinematic invariants
$$
s=(p_1+p_2)^2,  \qquad
t=(p_1-p_1')^2, \qquad
u=(p_1-p'_2)^2
$$
which are related as $s+t+u=4m^2$. The amplitude $T_{ij}^{i'j'}$ of elastic
quark-quark scattering contains color indices of both incoming ($i,j$) and
outgoing ($i',j'$) quarks. It depends on these invariants as well as on the
quark mass $m$ and infrared cutoff $\lambda$ (a fictitious gluon mass, for
instance) one has to introduce to regularize IR divergences. We will study
the asymptotic behavior of the amplitude of the near forward elastic
quark-quark scattering in the following region of parameters:
$$
s,\ m^2 \gg -t \gg \lambda^2 \gg \Lambda^2_{\QCD} .
$$
The amplitude $T_{ij}^{i'j'}$ is a function of the ratios of the kinematic
invariants. One could propose a lot of such ratios but, as we will show
below, only one possibility is realized:%
\footnote{This dependence is confirmed by calculations \ci{PT} of the
          scattering amplitude to the lowest orders of PT}
\be
T_{ij}^{i'j'}=T_{ij}^{i'j'}\lr{\frac{s}{m^2},\frac{t}{\lambda^2}}
\lab{fun}
\ee
Moreover, the $s-$dependence of the amplitude comes through the dependence
on the angle $\gamma$ between quark $4-$velocities $v_1=p_1/m$ and
$v_2=p_2/m$ defined in Minkowski space-time as
$$
(v_1\cdot v_2)=\cosh\gamma, \qqqquad
s=2m^2(1+\cosh\gamma)=4m^2\cosh^2(\gamma/2)\, .
$$
Let us go to the center of mass frame of the incoming quarks. In this frame,
quark momenta have the following light-cone components:
$$
p_1=(p_1^+,p_1^-,\vec p_1)
   =\frac{m}{\sqrt 2}(\e^{\gamma/2},\e^{-\gamma/2},\vec 0), \qquad
p_2=(p_2^+,p_2^-,\vec p_2)
   =\frac{m}{\sqrt 2}(\e^{-\gamma/2},\e^{\gamma/2},\vec 0)\, .
$$
In the limit $s\gg m^2$ the angle between quark velocities becomes large
$\gamma=\log\frac{s}{m^2} \gg 1$ and both quarks move close to the ``$+$''
and ``$-$'' light-cone directions. The total momentum transfer is
$q=p_1-p_1'$, and in the center of mass frame its components are
$
q^+=-q^-=\frac1{\sqrt 2}\frac{t}{\sqrt{s-4m^2}}
        =\CO\lr{{t}/{\sqrt s}}
$
and
$
{\vec q\,}^2={tu}/(s-4m^2)=\CO(t).
$
Thus, in the limit $s\gg -t$, we can neglect the longitudinal components
of the transferred momentum and put $q=(0^+,0^-,\vec q)$ with
$q^2\equiv t = -{\vec q\,}^2$.

Note, that the transferred momentum is much smaller than the energies of
incoming quarks. This suggests that the incoming quarks interact with each
other by exchanging soft gluons in the $t-$channel with the total momentum
$q$. This interaction gives rise to infrared divergences of the scattering
amplitude. It is a general feature of soft gluons \ci{IR} that their
contribution to the scattering amplitude is factorized into a universal
factor given by a Wilson line vacuum expectation value. The origin of this
property is the following. Interacting with soft gluons quark behaves as a
relativistic charged classical particle, and the only effect of its
interaction with soft gluons is the appearance of an additional phase in
the quark wave function. This phase, the so called eikonal phase, is defined
as a Wilson line $\CP\exp(i\int_C dx\cdot A(x))$ evaluated along the
classical trajectory $C$ of the quark. Applying the same approximation to
the near forward elastic quark-quark scattering we get two Wilson lines
corresponding to two incoming quarks. Note, that although quarks change
their velocities after scattering, we can neglect this difference in the
limit $-t \ll m^2$. It means that calculating the eikonal phases we may
consider incoming quarks as classical particles moving from $-\infty$ to
$+\infty$ with velocities $v_1$ and $v_2$. We combine two eikonal phases of
both incoming quarks and get the expression for the scattering amplitude
in the form
\be
T_{ij}^{i'j'} \lr{\frac{s}{m^2},\frac{q^2}{\lambda^2}}
=\sinh\gamma\int d^2z\, \e^{-i\vec z\cdot\vec q}
\langle 0|\CT\, W_+^{i'i}(0) W_-^{j'j}(z)|0\rangle\, ,
\qquad t=-{\vec q\,}^2
\lab{main}
\ee
Here, Wilson lines $W_+$ and $W_-$ are evaluated along infinite lines
in the direction of the quark velocities $v_1$ and $v_2$, respectively:
\be
W_+(0)=\CP\exp\left(
       i\int_{-\infty}^{\infty}d\alpha\ v_1\cdot A(v_1\alpha)\right)\, ,
\qquad
W_-(z)=\CP\exp\left(
       i\int_{-\infty}^{\infty}d\beta\ v_2\cdot A(v_2\beta+z)\right)\, ,
\lab{W-def}
\ee
and the integration paths are separated by the impact vector
$z=(0^+,0^-,\vec z)$ in the transverse direction as shown in fig.1.
\par\bigskip
\begin{center}
\unitlength=0.5mm
\linethickness{0.4pt}
\begin{picture}(95.00,85.00)
\put(20.00,20.00){\vector(1,1){50.00}}
\put(45.00,45.00){\vector(1,3){5.00}}
\put(54.00,56.00){\vector(-1,1){29.00}}
\put(56.00,54.00){\line(1,-1){19.00}}
\put(48.00,42.00){\makebox(0,0)[cc]{$0$}}
\put(53.00,63.00){\makebox(0,0)[cc]{$z$}}
\put(22.00,80.00){\makebox(0,0)[cc]{$i'$}}
\put(70.00,65.00){\makebox(0,0)[cc]{$j'$}}
\put(70.00,30.00){\makebox(0,0)[cc]{$i$}}
\put(26.00,18.00){\makebox(0,0)[cc]{$j$}}
\put(36.00,81.00){\makebox(0,0)[cc]{$v_1$}}
\put(61.00,70.00){\makebox(0,0)[cc]{$v_2$}}
\end{picture}

\parbox[t]{140mm}
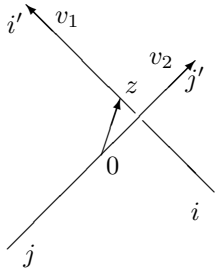
{Fig.~1: Wilson lines are evaluated along classical trajectories
of quarks with velocities $v_1$ and $v_2$ separated by the impact
vector $z=(0^+,0^-,\vec z)$ in the transverse direction. }
\end{center}
\noindent
Integration over two-dimensional impact vector $z$ was introduced to ensure
the total transferred momentum to be equal to $q=(0^+,0^-,\vec q)$. Indeed,
by expanding the Wilson lines in powers of gauge fields and using the
momentum representation for these fields one gets that expression \re{main},
firstly, reproduces the eikonal approximation for interaction vertices of
incoming quarks with soft gluons and, secondly, the total momentum of gluons
$k_{tot.}$ propagating in the $t-$channel is restricted by the following
$\delta-$functions:
$$
\delta(k_{tot.}\cdot v_1)\delta(k_{tot.}\cdot v_2)\delta(\vec k_{tot.}-\vec q)
=\frac{1}{\sinh \gamma}
\delta(k_{tot.}^+)\delta(k_{tot.}^-)\delta(\vec k_{tot.}-\vec q)
=\frac{1}{\sinh \gamma}\delta(k_{tot.}-q)
$$
where the first $\delta-$function comes from $W_+$, the second one
from $W_-$ and the last one from integration over $\vec z$. To
compensate the additional factor in the r.h.s. of this relation,
the same factor was introduced in the expression \re{main}.

\section{Properties of Wilson loops in QCD}

The calculation of vacuum expectation values of Wilson lines entering into
the expression \re{main} for the scattering amplitude is a hard problem
in PT.%
\footnote{The effective two-dimensional field theory proposed in \ci{rec}
     enables us to calculate this object only to the lowest orders of PT}
Let us consider first a special case of the elastic electron-electron
scattering in QED. One may use the expression \re{main} for the scattering
amplitude in this case, but the calculation of the Wilson lines is simplified
because for abelian gauge group a path-ordered exponent coincides with an
ordinary exponent. Moreover, since photons don't interact with each other,
the calculation of Wilson lines in QED is reduced to the integration of a
free photon propagator along the path of fig.1 and the result of calculation
is
\be
\bra{0} \CT W_+(0) W_-(z) \ket{0} =
\exp\lr{-\int dx^\mu\int dy^\nu D_{\mu\nu}(x-y)}
=\lr{\lambda^2 {\vec z\,}^2/4} {}^{i\frac{e^2}{4\pi}\coth\gamma}
\lab{W-QED}
\ee
where $\lambda$ is a fictitious photon mass. After substitution of this
relation into \re{main} and integration over impact vector we get
the expression for the scattering amplitude in QED \ci{QED}
$$
T\lr{\frac{s}{m^2},\frac{t}{\lambda^2}} = ie^2\frac{\cosh\gamma}{t}
\left(\frac{-t}{\lambda^2}\right)^{-i\frac{e^2}{4\pi}\coth\gamma}
\frac{\Gamma(1+i\frac{e^2}{4\pi}\coth\gamma)}
     {\Gamma(1-i\frac{e^2}{4\pi}\coth\gamma)}
$$
Of course, there are no reasons to expect that calculation of the Wilson
lines in QCD will be simple, but there are some features of the scattering
amplitude in QED which have their analogs in QCD. In particular, the
scattering amplitude obeys the condition \re{fun} and its $s-$dependence
comes from the dependence of Wilson lines on the angle $\gamma$ between
electron velocities. In the limit $s/m^2\to\infty$ or, equivalently
$\gamma\to\infty$, the expectation value \re{W-QED} doesn't depend on $s$
and, as a consequence, the scattering amplitud has the asymptotics
$T/T_{Born}\sim s^0$ which implies that photon is not reggeized in QED.
Here, $T_{Born}=\frac{ie^2}{t}\frac{s}{m^2}$ is the expression for the
amplitude in the Born approximation.

The asymptotics of the scattering amplitude \re{main} in QCD is defined by
properties of the line function equal to the vacuum expectation value of the
product of two Wilson lines:
\be
W^{i'j'}_{ij} \equiv \bra{0} \CT W_+^{i'i}(0) W_-^{j'j}(z)\ket{0}
\lab{W}
\ee
It depends on the color indices of incoming and outgoing quarks and is gauge
invariant. According to the definitions \re{W-def} and \re{W}, the line
function $W$ depends, in general, on the quark velocities $v_1$ and $v_2$, the
impact vector $z$ and infrared cutoff $\lambda$ which is introduced into
Feynman integrals to regularize IR divergences. Trying to form dimensionless
scalar invariants formed by these variables we find that, due to identities
$v_1^2=v_2^2=1$ and $(v_1z)=(v_2z)=0$, there is only one possibility
\be
W^{i'j'}_{ij}=W^{i'j'}_{ij}((v_1v_2),\lambda^2{\vec z\,}^2)
\equiv W^{i'j'}_{ij}(\gamma,\lambda^2{\vec z\,}^2)
\lab{W-fun}
\ee
After substitution into \re{main}, this relation implies the functional
dependence \re{fun} of the scattering amplitude. Thus, the $s-$dependence
of the scattering amplitude comes from the dependence of the line function
on the angle between quark trajectories in fig.1 while its $t-$dependence
is related to the dependence of the line function on the impact vector $z$.

To understand the $z-$dependence of line function, let us consider as an
example the one-loop calculation of $W$ in the Feynman gauge. Using the
definitions \re{W-def} and \re{W}, we get
\be
W_{1-loop}=I\otimes I+(t^a\otimes t^a)\frac{g^2}{4\pi^{D/2}}\Gamma(D/2-1)
\lambda^{4-D}
\int_{-\infty}^{\infty}d\alpha\int_{-\infty}^{\infty}d\beta
\frac{(v_1v_2)}{[-(v_1\alpha-v_2\beta)^2+{\vec z\,}^2+i0]^{D/2-1}}
\lab{W-one}
\ee
where a direct product of the gauge generators defined in the fundamental
representation takes care of the color indices of the quarks. In this
expression, gluon is attached to both Wilson lines at points $v_1\alpha$
and $v_2\beta+z$ and we integrate the gluon propagator
$v_1^\mu v_2^\nu D_{\mu\nu}(v_1\alpha-v_2\beta-z)$
over positions of these points.
To regularize IR divergences we introduced the dimensional regularization
with $D=4+2\varepsilon,$ ($\varepsilon > 0$) and $\lambda$ being the IR
renormalization parameter. There is another contribution to $W_{1-loop}$
corresponding to the case when gluon is attached by both ends to one of
the Wilson lines. A careful treatment shows that this contribution
vanishes. Performing integration over parameters $\alpha$ and $\beta$ we get
\be
W_{1-loop}(\gamma,\lambda^2{\vec z\,}^2)
=I\otimes I+(t^a\otimes t^a)\alpi (-i\pi\coth\gamma)
\Gamma(\varepsilon)
(\pi\lambda^2{\vec z\,}^2)^{-\varepsilon}
\lab{W-z}
\ee
The integral over $\alpha$ and $\beta$ in \re{W-one} has an infrared
divergence coming from large $\alpha$ and $\beta$. In the dimensional
regularization, this divergence appears in $W_{1-loop}$ as a pole in $(D-4)$
with the renormalization parameter $\lambda$ having a sense of an IR cutoff.
This result may seem strange because, after substitution of the one-loop
expression for $W$ into \re{main}, we should reproduce the expression for
the scattering amplitude in the Born approximation
$T_{Born}=\frac{ig^2}{t}\frac{s}{m^2}(t^a\otimes t^a)$
corresponding to one-gluon exchange in the $t-$channel. Indeed, performing a
Fourier transformation of $W_{1-loop}$ in \re{main} we find that IR divergence
disappears, and the one-loop expression $T_{Born}$ for the scattering
amplitude is reproduced.

The approach we propose for the calculation of Wilson lines vacuum expectation
value is based on the following observation. The one-loop expression \re{W-z}
for the line function is divergent for $\vec z=0$. This divergence has an
ultraviolet origin
because it comes from integration over small $\alpha$ and $\beta$ in
\re{W-one}, that is from gluons propagating at short distances
($v_1\alpha-v_2\beta$) between quark trajectories. One should notice that
the ultraviolet (UV) divergences of Wilson lines at $z=0$ have nothing
to do with the ``conventional'' ultraviolet singularities in QCD. For
$z=0$, the integration paths of Wilson lines of fig.1 cross each other
at point $0$ and UV divergences appear when one integrates gluon propagators
along integration path at the vicinity of the cross point. It is not for
the first time when one encounters these very specific divergences of
Wilson lines. The same divergences, the so called cross divergences, were
observed more than 10 years ago \ci{cross} when attempts have been made to
reformulate the dynamics of gauge fields in terms of string operators.

We note that the cross divergence has not appeared in $W_{1-loop}$
for nonzero $z$ because, as it enters into the integral \re{W-one},
nonzero ${\vec z\,}^2$ regularizes gluon propagator at short distances.
Thus, ${\vec z\,}^2$
has a meaning of an UV cutoff for the one-loop line function $W$ and
the $z-$dependence of the original line function
$W(\gamma,\lambda^2{\vec z\,}^2)$
(or, equivalently, the $t-$dependence of the scattering amplitude) is in
one-to-one correspondence with the dependence of the same Wilson line but
with $z=0$ on the UV cutoff which we have to introduce to regularize the
cross singularities. The general reason for this property to be valid to
all orders of PT is the following. As follows from \re{W-fun}, the line
function \re{W} depends only on two dimensionfull arguments: $z^2$ and
$\lambda^2$ while quark velocities are dimensionless. It implies that
Feynman integrals over gluon momenta which one gets by calculating
vacuum averaged Wilson lines in the perturbation theory contain only two
momentum scales: $\lambda^2$
and $1/z^2$. The fact that the first scale cuts gluon momenta from below
implies that the second scale cuts gluon momenta from above. For $z=0$,
this second scale goes to infinity and line function becomes UV divergent.
For ${\vec z\,}^2\neq 0$, the line function is finite but the cross
singularities manifest themselves in the singular dependence of $W$ on $z$
as ${\vec z\,}^2\to 0$.
The classical trajectories of quarks in fig.1 are shifted by vector $z$ in
the transverse direction, and we may consider this shift as one of possible
ways to regularize the cross singularities which line function $W$ will have
for $z=0$. This simple observation suggests the following way for calculation
of the $z-$dependence of the Wilson lines:
first, put $z=0$ in the definition \re{W} of line function and introduce the
      regularization of cross singularities of $W$;
second, renormalize cross singularities and identify UV cutoff with impact
      vector as $\mu^2=1/{\vec z\,}^2$.

Thus, the $z-$dependence of the line function $W$ and, as consequence, the
$t-$asymptotics of the scattering amplitude are controlled by the
renormalization properties of the cross singularities of Wilson loops.

\section{Renormalization of Wilson loops}
The line function $W^{i'j'}_{ij}$ has color indices of incoming and outgoing
quarks. Since quarks are defined in the fundamental representation of the
gauge group $SU(N)$ one can decompose $W$ into two invariant tensors
$\delta_{ii'}\delta_{jj'}$ and $\delta_{ij'}\delta_{ji'}$:
\be
W^{i'j'}_{ij}=\frac{N W_1^{(b)}-W_1^{(a)}}{N(N^2-1)}\delta_{ii'}\delta_{jj'}
           +\frac{N W_1^{(a)}-W_1^{(b)}}{N(N^2-1)}\delta_{ij'}\delta_{ji'}\,.
\lab{dec}
\ee
Projections of $W$ onto these tensors define two gauge invariant Wilson loops
$W_1^{(a)}=\delta_{ij'}\delta_{ji'}W^{i'j'}_{ij}$ and
$W_1^{(b)}=\delta_{ii'}\delta_{jj'}W^{i'j'}_{ij}$ which are evaluated along
the integration paths of fig. 2 (a) and (b), respectively, closed at infinity.
According to the approach formulated at the end of the previous section, we
put $z=0$ and consider renormalization properties of these Wilson loops having
a cross point.
\par\bigskip
\begin{center}
\unitlength=0.5mm
\linethickness{0.4pt}
\begin{picture}(205.00,70.00)
\put(20.00,20.00){\vector(1,1){50.00}}
\put(70.00,20.00){\line(-1,1){24.00}}
\put(44.00,46.00){\vector(-1,1){24.00}}
\put(130.00,20.00){\vector(1,1){50.00}}
\put(180.00,20.00){\line(-1,1){24.00}}
\put(154.00,46.00){\vector(-1,1){24.00}}
\put(45.00,39.00){\makebox(0,0)[cc]{$0$}}
\put(155.00,39.00){\makebox(0,0)[cc]{$0$}}
\put(45.00,2.00){\makebox(0,0)[cc]{(a)}}
\put(155.00,1.00){\makebox(0,0)[cc]{(b)}}
\bezier{25}(20.00,70.00)(0.00,45.00)(20.00,20.00)
\bezier{25}(70.00,69.00)(90.00,45.00)(70.00,20.00)
\bezier{68}(130.00,70.00)(95.00,-5.00)(180.00,20.00)
\bezier{68}(180.00,70.00)(215.00,-5.00)(130.00,20.00)
\end{picture}
\par\bigskip\par
\parbox[t]{140mm}
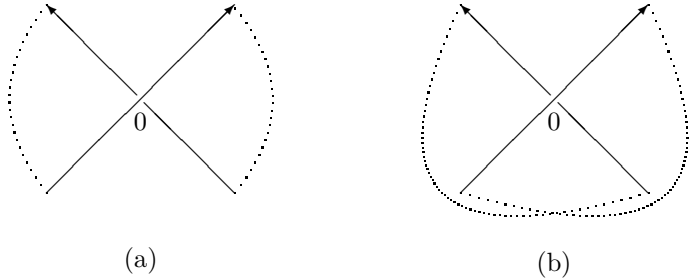
{Fig.~2: Integration paths entering into the definition of the
``physical'' Wilson loops $W_1^{(a)}$ and $W_1^{(b)}$, respectively.
Dotted line indicates the way in which these paths are closed at infinity.}
\end{center}
\noindent
It turns out \ci{cross} that the renormalization properties of Wilson loops
with a cross point depend on the representation of the gauge group in which
Wilson loop is defined. In the case of the quark-quark scattering, the Wilson
loops are defined in the quark representation but considering a near forward
gluon-gluon scattering in an analogous manner we will get the same expressions
for the scattering amplitude but with Wilson lines defined at the
{\it adjoint\/} representation of the gauge group. It is exactly the way in
which we will find the difference between the asymptotics of the amplitude of
quark-quark and gluon-gluon scattering.

For Wilson loops $W_1^{(a)}$ and $W_1^{(b)}$ defined in the fundamental
representation of the $SU(N)$ group the renormalization properties can
be formulated as follows \ci{cross}. The Wilson loop $W_1^{(a)}$ of fig.2(a)
is mixed under renormalization with the Wilson loop $W_2^{(a)}$ of fig.3(a),
and the Wilson loop $W_1^{(b)}$ of fig.2(b) is mixed with the Wilson loop
$W_2^{(b)}$ of fig.3(b).
\begin{center}
\unitlength=0.5mm
\linethickness{0.4pt}
\begin{picture}(215.00,70.00)
\put(20.00,20.00){\line(1,1){25.00}}
\put(45.00,45.00){\vector(-1,1){25.00}}
\put(71.00,20.00){\line(-1,1){25.00}}
\put(46.00,45.00){\vector(1,1){25.00}}
\put(46.00,39.00){\makebox(0,0)[cc]{$0$}}
\put(46.00,2.00){\makebox(0,0)[cc]{(a)}}
\put(129.00,20.00){\line(1,1){25.00}}
\put(154.00,45.00){\vector(-1,1){25.00}}
\put(180.00,20.00){\line(-1,1){25.00}}
\put(155.00,45.00){\vector(1,1){25.00}}
\put(155.00,39.00){\makebox(0,0)[cc]{$0$}}
\put(155.00,2.00){\makebox(0,0)[cc]{(b)}}
\bezier{26}(20.00,70.00)(-1.00,45.00)(20.00,20.00)
\bezier{25}(71.00,70.00)(91.00,45.00)(71.00,20.00)
\bezier{70}(129.00,70.00)(93.00,-7.00)(180.00,20.00)
\bezier{70}(180.00,70.00)(215.00,-7.00)(129.00,20.00)
\end{picture}
\par\bigskip\par
\parbox[t]{140mm}
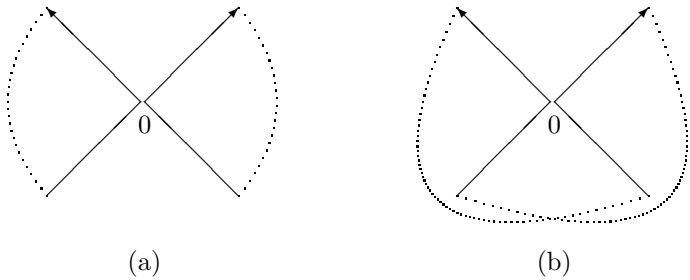
{Fig.~3: Wilson loops $W_2^{(a)}$ and $W_2^{(b)}$
which are mixed under renormalization with the
``physical'' Wilson loops of Fig.~2(a) and (b) respectively.}
\end{center}
\noindent
Both integration paths in fig.3(a) and 3(b) are
different from the original paths of fig.2(a) and 2(b) only at the vicinity
of the cross point. It is natural to introduce the following doublets:
$$
W^{(a)}=\bin{W_1^{(a)}}{W_2^{(a)}} ,
\qqqquad
W^{(b)}=\bin{W_1^{(b)}}{W_2^{(b)}}
$$
which don't mix with each other under renormalization.%
\footnote{Notice that the Wilson loops $W_2^{(a)}$ and $W_2^{(b)}$
          were introduced into consideration by the renormalization procedure
          and in contract with $W_1^{(a)}$ and $W_1^{(b)}$ they don't have any
          physical meaning}
Then both doublets obey the same renormalization group equation \ci{cross}
\be
\lr{ \mu\partder{\mu} + \beta(g)\partder{g} } W(\gamma,\lambda^2/\mu^2)
=-\Gamma_{cross}(\gamma,g) W(\gamma,\lambda^2/\mu^2)
\lab{W-RG}
\ee
where $\mu$ is the renormalization parameter and $\lambda$ is an IR cutoff.
A new object appears in the r.h.s. of this equation, a $2\times 2$ matrix
of anomalous dimensions $\Gamma_{cross}(\gamma,g)$. This matrix is gauge
invariant and it depends only on the coupling constant and the angle between
quark velocities at the cross point. Following \ci{cross} we calculate the
one-loop expression for this matrix in the fundamental representation of
the $SU(N)$ gauge group
\be
\Gamma_{cross}(\gamma,g)=\alpi\Gamma(\gamma)+\CO(\as^2),
\lab{cr}
\ee
where $2\times 2$ matrix $\Gamma(\gamma)$ depends only on the angle $\gamma$
and is given by
$$
\Gamma(\gamma)=\lr{ \begin{array}{cc}
  -\frac{i\pi}{N}\coth\gamma   &    i\pi \coth\gamma
\\
-\gamma\coth\gamma+1+i\pi\coth\gamma & N(\gamma\coth\gamma-1)
                                       -\frac{i\pi}{N}\coth\gamma
\end{array} }
$$
The RG equation \re{W-RG} implies that the $s-$dependence of Wilson loops is
defined by the properties of this matrix. The one-loop expression \re{cr}
is valid for an arbitrary values of $\gamma$. In the limit $s\gg m^2$, the
matrix $\Gamma(\gamma)$ has the following asymptotics:
$$
\Gamma(\gamma)=\lr{ \begin{array}{cc}
  -\frac{i\pi}{N} & i\pi
\\
  -\log\frac{s}{m^2}+1+i\pi & N(\log\frac{s}{m^2}-1)-\frac{i\pi}{N}
\end{array} }
$$
where we neglected terms vanishing as $s/m^2\to\infty$. To check the RG
equation \re{W-RG} we perform the one-loop calculation of Wilson loops
\ba
W_1^{(a)}(\gamma,\lambda^2/\mu^2)
&=&N\left(1-\frac{\alpha_s}{\pi}C_F\cdot 2i\pi\coth(\gamma)
\log\frac{\mu}{\lambda}\right)
\nonumber \\
W_2^{(a)}(\gamma,\lambda^2/\mu^2)
&=&N^2\left(1-\frac{\alpha_s}{\pi}C_F\cdot
2(\gamma\coth(\gamma)-1)
\log\frac{\mu}{\lambda}\right)
\nonumber \\
W_1^{(b)}(\gamma,\lambda^2/\mu^2)
&=&N^2\left(1-\frac{\alpha_s}{\pi}\cdot 0
\log\frac{\mu}{\lambda}\right)
\nonumber \\
W_2^{(b)}(\gamma,\lambda^2/\mu^2)
&=&N\left(1-\frac{\alpha_s}{\pi}C_F\cdot 2i\pi\coth(\gamma)
\log\frac{\mu}{\lambda}\right)
\ea
with $C_F=\frac{N^2-1}{2N}$
and verify that they do satisfy the RG equation \re{W-RG}. Note, that these
Wilson loops have different normalization at the lowest order. Moreover,
at $\mu=\lambda$ the one-loop corrections disappear
\be
W^{(a)}(\gamma,\lambda^2/\mu^2)|_{\mu=\lambda}
=\bin{N}{N^2} ,
\qqqquad
W^{(b)}(\gamma,\lambda^2/\mu^2)|_{\mu=\lambda}
=\bin{N^2}{N}
\lab{W-b.c.}
\ee
which means that for UV and IR cutoff for gluon momenta equal to each
other there is no phase space for gluons.

\section{Solving the RG equation}

By solving the RG equation \re{W-RG} with the boundary conditions \re{W-b.c.}
we could find the ``physical'' Wilson loops $W_1^{(a)}$ and $W_1^{(b)}$.
Different Wilson loops of figs.2 and 3 satisfy the same equation \re{W-RG}
and the reason why we get four different expressions for them lies in the
different boundary conditions \re{W-b.c.} we impose on the solutions of
\re{W-RG}. The general solution of the RG equation \re{W-RG} with the
boundary conditions \re{W-b.c.} for the doublet $W^{(a)}$ is given by
$\CT-$ordered exponent
\be
W^{(a)}(\gamma,\lambda^2/\mu^2)
=\CT \exp\lr{-\int_\lambda^\mu\frac{d\tau}{\tau}
              \Gamma_{cross}(\gamma,g(\tau))}\bin{N}{N^2}\, ,
\lab{W-I}
\ee
and for the second doublet we get analogous expression
\be
W^{(b)}(\gamma,\lambda^2/\mu^2)
=\CT \exp\lr{-\int_\lambda^\mu\frac{d\tau}{\tau}
              \Gamma_{cross}(\gamma,g(\tau))}\bin{N^2}{N}\, .
\lab{W-II}
\ee
The matrices $\Gamma_{cross}(\gamma,g(\tau))$ don't commute with each other
for different values of $\tau$ and it makes very difficult to evaluate the
$\CT-$exponents.%
\footnote{We may follow another route and eliminate one of the components of
          the doublet $W$ from the system of equation \re{W-RG} to get the
          equation for the second component. The resulting second order
          differential equation is known as the Riccatti equation}
To find the scattering amplitude we extract the upper components of the
doublets \re{W-I} and \re{W-II} and get two ``physical'' Wilson loops
$W_1^{(a)}$ and $W_1^{(b)}$,
evaluate the Wilson lines $W_{ij}^{i'j'}$ using
\re{dec}, identify the renormalization parameter as $\mu^2=1/{\vec z\,}^2$ and
substitute $W_{ij}^{i'j'}$ into \re{main}:
\be
T_{ij}^{i'j'} \lr{\frac{s}{m^2},\frac{t}{\lambda^2}}
=\sinh\gamma\int d^2z\, \e^{-i\vec z\cdot\vec q}
\lr{ A_{11}(\gamma,z^2\lambda^2)\delta_{ii'}\delta_{jj'}
+A_{12}(\gamma,z^2\lambda^2)\delta_{ij'}\delta_{ji'}}\, ,
\quad t=-{\vec q\,}^2
\lab{res0}
\ee
where $A_{11}$ and $A_{12}$ are elements of the $2\times 2$ matrix
$A\lr{\gamma,z^2\lambda^2}=\CT \exp\lr{-\int_\lambda^{1/z}
\frac{d\tau}{\tau}\Gamma_{cross}(\gamma,g(\tau))}$.

The simplification occurs when we substitute the one-loop expression
\re{cr} for the matrix $\Gamma_{cross}$ into \re{W-I} and \re{W-II}.
One-loop matrices $\Gamma_{cross}(\gamma,g(\tau))$ commute with each
other for different $\tau$ and we can omit $\CT-$ordering in the
definition of the matrix $A$. After diagonalization of the one-loop
matrix $\Gamma(\gamma)$ by a proper unitary transformation we get
$$
A\lr{\gamma,{\lambda^2}{z^2}}=
\frac{\Gamma_+-\Gamma(\gamma)}{\Gamma_+-\Gamma_-}\
\exp\lr{ -\Gamma_- \int_\lambda^{1/z}\frac{d\tau}{\tau}\al{\tau} }
+
\frac{\Gamma_--\Gamma(\gamma)}{\Gamma_--\Gamma_+}\
\exp\lr{ -\Gamma_+ \int_\lambda^{1/z}\frac{d\tau}{\tau}\al{\tau} }
$$
Here,
$\Gamma_\pm$ are
the eigenvalues of the matrix
$\Gamma(\gamma)$ which satisfy the following characteristic equation
$$
\Gamma_\pm^2
-\Gamma_\pm \lr{ N(\gamma\coth\gamma-1)-\frac{2i\pi}{N}\coth\gamma }
+\pi^2\coth^2\gamma=0
$$
Finally,
we find the following expression for
the scattering amplitude \re{res0}
\be
T_{ij}^{i'j'}=\delta_{i'i}\delta_{j'j}\ T^{(0)}
             +t^a_{i'i} t^a_{j'j}\ T^{(8)}
\lab{res1}
\ee
where two invariant amplitudes corresponding to the exchange in the
$t-$channel by states with the quantum numbers of vacuum and gluon
are given by
\be
T^{(0)}=-\frac{\sinh\gamma}{t}\frac{\Gamma_+\Gamma_-}{\Gamma_+-\Gamma_-}
(T_+-T_-),
\qquad
T^{(8)}=2i\pi\frac{\cosh\gamma}{t}\frac{\Gamma_+T_+-\Gamma_-T_-}
{\Gamma_+-\Gamma_-}
\lab{res2}
\ee
Here the notation was introduced for the Fourier transforms
$$
T_\pm=\frac{t}{\Gamma_\pm}\int d^2z\ \e^{-i\vec z\cdot\vec q}
\exp\lr{ -\Gamma_\pm\int_\lambda^{1/z}\frac{d\tau}{\tau} \al{\tau} }
     =\frac{t}{\Gamma_\pm}\int d^2z\ \e^{-i\vec z\cdot\vec q}
\left( \frac{\log 1/z^2\Lambda_{\QCD}^2}{\log\lambda^2/\Lambda_{\QCD}^2}
\right)^{-2\Gamma_\pm/\beta_0}
$$
To the lowest order of PT we have $T_\pm=2\as$ and after substitution of
these values into \re{res2} we recover one-loop expression for the scattering
amplitude: $T^{(0)}=0$ and $T^{(8)}=\frac{ig^2}{t}\frac{s}{m^2}$. To get the
expression for $T_\pm$ in the leading $\log t-$approximation we freeze the
argument of the coupling constant and after integration get
\be
T_\pm\simeq 2\as
\exp\lr{-\frac{\as}{2\pi}\Gamma_\pm\log\frac{-t}{\lambda^2}} \, .
\lab{lla}
\ee
Let us consider the properties of the obtained expressions \re{res1} in
the limit
of high energies $s \gg m^2$. Solving the characteristic equation
we find that in the large $s$ limit both eigenvalues have positive
real part (Re $\Gamma_\pm >0$). Moreover, one of the eigenvalues of matrix
$\Gamma$ is much larger than the second one:
\baa
\Gamma_+ &=& N\log\frac{s}{m^2}- N-\frac{2i\pi}{N}
-\pi^2\frac{N^2-1}{N^3}\log^{-1}\frac{s}{m^2}
+ \CO\lr{\log^{-2}\frac{s}{m^2}},
\qquad
\\
\Gamma_- &=&
\pi^2\frac{N^2-1}{N^3}\log^{-1}\frac{s}{m^2}
+ \CO\lr{\log^{-2}\frac{s}{m^2}}
\eaa
Together with \re{res2} this property of the matrix $\Gamma_{cross}$ implies
that the amplitudes of singlet and octet exchanges have the following
high-energy behavior
$$
T^{(0)}\propto \Gamma_- T_+ - \Gamma_- T_-\, ,
\qquad
T^{(8)}\propto T_+-\frac{\Gamma_-}{\Gamma_+}T_- \,.
$$
The functions $T_+$ and $T_-$ have a Regge like behavior and for
$\Gamma_+ \gg \Gamma_-$ we have $T_+\ll T_-$. Moreover, the contribution of
$T_-$ to the amplitude of the octet exchange is suppressed by the factor
$1/\Gamma_+=\CO(\log^{-1} s/m^2)$ as compared to that to the amplitude of the
singlet exchange. Thus, the high-energy asymptotic behavior of the scattering
amplitude \re{res1} is dominated by the contribution of $T_-$ to the singlet
exchange amplitude $T^{(0)}$.

To get the expression for the scattering amplitude \re{res1} in the leading
logarithms in $s$ and $t$ we substitute $\Gamma_+=N\log s/m^2$ and $\Gamma_-=0$
into \re{res2} and \re{lla}. The result has the standard reggeized form
\ci{PT}:
$
T=T_{Born} (s/m^2)^{\alpha(t)}
$
with the Regge trajectory $\alpha(t)
=-\frac{\alpha_s}{2\pi}N\log\frac{-t}{\lambda^2}
=-\frac{\alpha_s}{4\pi^2}N\int\frac{{\vec q\,}^2\, d^2 k}
{{\vec k\,}^2(\vec q-\vec k)^2}$.

\section{Conclusions}

In this paper we considered the elastic quark-quark scattering at high energy
$s$ and fixed transferred momentum $t$. Performing factorization of soft gluon
emissions, we found the expression for the scattering amplitude \re{main} as a
Fourier transformed vacuum expectation value of a product of two Wilson lines
defined in the quark representation of the $SU(N)$ gauge group and evaluated
along the classical trajectories of quarks with velocities $v_1$ and $v_2$.
We calculated these Wilson lines using the fact that in the limit of zero
impact vector $z$, when quark trajectories intersect each other, Wilson lines
get very specific UV cross divergences. There is a close relation between the
renormalization properties of cross divergences and the $z-$dependence of
Wilson lines or, equivalently, the $t-$dependence of the scattering amplitude.
Using this relation we found the scattering amplitude \re{res0} which depends
on the $2\times2$ matrix of anomalous dimensions $\Gamma_{cross}(\gamma,g)$ of
Wilson loops. Performing the one-loop calculation of this matrix we found that
these are the properties of the matrix $\Gamma_{cross}(\gamma,g)$ which
determine the asymptotics of the scattering amplitude \re{res1}. Being a
function of the angle $\gamma$ between quark velocities, the matrix
$\Gamma_{cross}(\gamma,g)$ depends on the energy $s$ and one has a regular
way \ci{cross} for its calculation in higher orders of PT. The important
question arises: whether higher order corrections will change the asymptotics
of the one-loop matrix \re{cr} and the properties of the eigenvalues
$\Gamma_\pm$ which were important for us to get the asymptotics in \re{res1}.
This will be the subject of a separate paper.

We may generalize the consideration to study the high-energy elastic
quark-antiquark and gluon-gluon scattering. In both cases, the scattering
amplitude is given in a close analogy with \re{main} by the vacuum
expectation value of two Wilson lines coming as eikonal phases of incoming
particles. It is important to recognize that Wilson lines carry the same
representation at which these particles are defined. For quark-antiquark
scattering, the Wilson line coming from antiquark is evaluated along its
velocity $v_2$, say, with gauge generators in the antiquark representation.
We notice that in this situation an antiquark can be treated as a quark
moving backwards in time with velocity $-v_2$ and all previous results for
the quark-quark scattering amplitude can be applied provided that we replace
one of the quark velocities as $v_2\to -v_2$. In terms of the angle $\gamma$
between quark velocities this transformation corresponds to the replacement
$\gamma \to i\pi-\gamma$ in $\Gamma_{cross}(\gamma,g)$.
For gluon-gluon scattering both Wilson lines are defined in the adjoint
representation of the $SU(N)$ gauge group. Following the analysis of
Sect.3, we relate the asymptotics of the gluon-gluon scattering amplitude
to the renormalization properties of Wilson loops of fig.2. As it was
stressed before, these properties are different for the fundamental
(quark) and the adjoint (gluon) representations of the $SU(N)$ gauge group.
One may find a novel feature of cross singularities in the adjoint
representation by considering the one-loop expression \re{W-z} for the Wilson
lines in the simplest case of the $SU(2)$ gauge group. Simplifying the direct
product of the gauge generators of the $SU(2)$ group we get the following
color flow:
$
t^a_{i'i}t^a_{j'j}=\frac12\delta_{ij'}\delta_{i'j}
                   -\frac14\delta_{i'i}\delta_{j'j}
$
in the fundamental representation and
$
t^a_{i'i}t^a_{j'j}=\delta_{ij'}\delta_{i'j}
                   -\delta_{ij}\delta_{i'j'}
$
in the adjoint representation. Different color flows in these two
representations imply that Wilson loop of fig.2(a) defined in the fundamental
representation of $SU(2)$ is mixed under renormalization with the Wilson loop
of fig.3(a) while in the adjoint representation there is an additional Wilson
loop of fig.4 which is mixed with both of them.
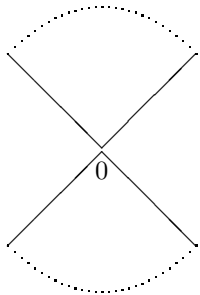
\begin{center}
\unitlength=0.5mm
\linethickness{0.4pt}
\begin{picture}(90.00,95.00)
\put(45.00,39.00){\makebox(0,0)[cc]{$0$}}
\bezier{28}(20.00,19.00)(45.00,-6.00)(70.00,19.00)
\bezier{28}(20.00,70.00)(45.00,95.00)(70.00,70.00)
\put(20.00,70.00){\line(1,-1){25.00}}
\put(45.00,45.00){\line(1,1){25.00}}
\put(20.00,19.00){\line(1,1){25.00}}
\put(45.00,44.00){\line(1,-1){25.00}}
\end{picture}
\par\bigskip\par
\parbox[t]{140mm}
{Fig.~4: Wilson loop which is mixed under renormalization with the
Wilson loops of Fig.~2(a) and 3(a) defined at the adjoint representation
of the $SU(2)$ group.}
\end{center}
\noindent
Coming back to the $SU(N)$
group, we note that the renormalization of cross singularities was performed
\ci{cross} only in the fundamental representation and the generalization to
other representations is an open question.

There have been a number of attempts in the 80's to describe the confining
dynamics of QCD in terms of Wilson loops. One of the problems which has
arisen on this way was \ci{pro,cross} that Wilson loop possesses extra UV
singularities when the integration path has end point, contains cusp or
crosses itself at some point. The renormalization of these singularities
introduces into consideration new anomalous dimensions of Wilson loops:
$\Gamma_{end}(g)$, $\Gamma_{cusp}(\gamma,g)$ and $\Gamma_{cross}(\gamma,g)$,
respectively. It turns out \ci{cusp} that these anomalous dimensions which
appear as undesired features of Wilson loops are of the most importance in
perturbative QCD when one studies effects of soft gluons. Namely, the
infrared behavior of quark and gluon propagators is controlled by
$\Gamma_{end}$. The asymptotics of the Sudakov form factor, the structure
functions of deep inelastic scattering for $x\to 1$, large perturbative
corrections to the Drell-Yan cross section are determined by $\Gamma_{cusp}$,
the same function was introduced in the heavy quark effective field theory
as the velocity dependent anomalous dimension. Finally, as it was shown in
this paper, it is $\Gamma_{cross}$, which governs the high energy behavior
of the parton-parton scattering amplitudes.

\bigskip\par\noindent{\Large {\bf Acknowledgements}}\par\bigskip\par

\noindent
I am most grateful to G.~Sterman for numerous stimulating discussions.
I would like to thank J.~Collins, G.~Marchesini and A.~Radyushkin for
helpful conversations.
This work is supported by the National Science Foundation
under grant PHY9309888 and by the Texas National Research Laboratory.

\bb{99}
\bi{book1}
      L.N. Lipatov, in {\it ``Perturbative QCD''\/}, ed. A.H. Mueller
      (World Scientific, Singapore, 1989) and references therein.
\bi{book2}
      H. Cheng and T.T. Wu, {\it ``Expanding Protons: Scattering at
      High Energies''\/}, (MIT Press, Cambridge, Massachusetts, 1987).
\bi{PT}
      H.T.Nieh and Y.-P.Yao, Phys. Rev. Lett. 32 (1974) 1074;
                             Phys. Rev. D13 (1976) 1082;
\\    B.M.McCoy and T.T.Wu, Phys. Rev. Lett. 35 (1975) 604;
                            Phys. Rev. D12 (1975) 3257;
\\    L.Tyburski, Phys. Rev. D13 (1976) 1107;
\\    C.Y.Lo and H.Cheng, Phys. Rev. D13 (1976) 1131.
\bi{QED}
        H. Cheng and T.T. Wu, Phys. Rev. Lett. 22 (1969) 666;
\\      H. Abarbanel and C. Itzykson,  Phys. Rev. Lett. 23 (1969) 53.
\bi{2-dim}
      L.N. Lipatov, Nucl. Phys. B365 (1991) 641;
                    Phys. Lett. B309 (1993) 394;
      {\it ``High energy asymptotics of multi-color QCD and exactly
      solvable lattice models''\/}, Padova preprint, DFPD/93/TH/70,
      October 1993.
\bi{i-pi}
      L.N. Lipatov, Nucl. Phys. B309 (1988) 379.
\bi{rec}
       H.Verlinde and E.Verlinde, {\it ``QCD at high energies and
       two-dimensional field theory''\/}, Princeton Univ. preprint,
       PUPT-1319, September 93.
\bi{evol}
        M.G.Sotiropoulos and G.Sterman, {\it ``Color exchange
        in near-forward hard elastic scattering''\/}, Stony Brook preprint,
        ITP--SB--93--59, October 93.
\bi{IR}
        A.Bassetto, M.Ciafaloni and G.Marchesini,
        Phys. Rept. 100 (1983) 201;
\\      J.C.Collins, D.E.Soper and G.Sterman, in {\it ``Perturbative QCD''\/},
        ed. A.H. Mueller (World Scientific, Singapore, 1989).
\bi{cross}
        R.A.Brandt, F.Neri and M.-A.Sato, Phys. Rev. D24 (1981) 879;
\\      R.A.Brandt, A.Gocksch, M.-A.Sato and F.Neri,
        Phys. Rev. D26 (1982) 3611.
\bi{pro}
        A.M.Polyakov, Nucl. Phys. B164 (1980) 171.
\bi{cusp}
        G.P.Korchemsky and A.V.Radyushkin,
        Sov. J. Nucl. Phys. 44 (1986) 145; 45 (1987) 127; 910;
        Phys. Lett. 171B (1986) 459; Phys. Lett. B279 (1992) 359;
\\      G.P.Korchemsky, Phys. Lett. 217B (1989) 330; 220B (1989) 629;
        Mod. Phys. Lett. A4 (1989) 1257.
\eb
\end{document}